\documentstyle[ApJ,times,PSfig]{article}

\def\beq{\begin{equation}} \def\eeq{\end{equation}}
\def\ie{i.e$.$~}  \def\etal{et al$.$~}
\def\eq{eq$.$~} \def\eqs{eqs$.$~} \def\dd{{\rm d}}
\def\epsel{\varepsilon_e} \def\epsmax{\epsilon}
\def\epsmag{\varepsilon_B} \def\adindex{\hat{\gamma}}
\def\cm3{\;{\rm cm^{-3}}} \def\deg{^{\rm o}}
\def\Meszaros{M\'esz\'aros~}

\def\simg{\mathrel{%
      \rlap{\raise 0.511ex \hbox{$>$}}{\lower 0.511ex \hbox{$\sim$}}}}
\def\siml{\mathrel{%
      \rlap{\raise 0.511ex \hbox{$<$}}{\lower 0.511ex \hbox{$\sim$}}}}

\begin{document}

\title{Jet Energy and Other Parameters for the Afterglows of Gamma-Ray Bursts 980703, 
       990123, 990510, and 991216 Determined from Modeling of Multi-Frequency Data}
\author{A. Panaitescu}
\affil{Dept. of Astrophysical Sciences, Princeton University, Princeton, NJ 08544}
\and
\author{P. Kumar}
\affil{Institute for Advanced Study, Olden Lane, Princeton, NJ 08540}

\begin{abstract}

 We model the radio, optical, and $X$-ray emission for the afterglows of GRB 980703, 
990123, 990510, and 991216, within the framework of relativistic jets, to determine 
their physical parameters. The models that yield acceptable fits to the data have jet 
energies mostly between $10^{50}$ to $10^{51}$ erg and initial opening angles between 
$1\deg$ and $4\deg$. The external medium density is uncertain by at least one order of 
magnitude in each case, being around $10^{-3}\cm3$ for GRB 980703 and 990123,  $\sim
10^{-1}\cm3$ for GRB 990510, and $\sim 3 \cm3$ for GRB 991216. 
If the jets are uniform (\ie there are no angular gradients of the energy per solid 
angle) then the 20 keV -- 1 MeV radiative efficiency during the GRB phase must have 
been at least 2-3\% for GRB 990510, 20\% for GRB 990123, and 30\% for GRB 991216. 

\end{abstract}


\section{Introduction}

 There are two basic quantities one needs to try understanding the nature of 
any astronomical source -- the distance and the energy of the source. For the 
long duration GRBs, lasting more than 10 seconds, the former is well established 
to be cosmological. However, the energy associated with the GRBs remains uncertain.

 The efficiency of producing $\gamma$-ray emission in the generally accepted 
internal shock model (\Meszaros \& Rees 1994), which can explain the observed 
temporal variability, is less than a few percent (Kumar 1999, Lazzati, Ghisellini
\& Celotti 1999, Panaitescu, Spada \& \Meszaros 1999, see also Beloborodov 2000). 
This suggests that the total energy in the explosion is larger than the energy 
observed in $\gamma$-rays by a factor of 20--50. 

 The goal of this work is to infer the physical parameters of the ejecta, 
including their energy and the external medium density, from modeling of radio, 
optical and $X$-ray data for GRB afterglows with known redshifts. The modeling is 
carried out in the framework of collimated ejecta interacting with an isotropic 
external medium. The model is described in \S\ref{model} and the results of the
numerical calculations for individual afterglows are given in \S\ref{aglows}.

\section{Description of the Model}
\label{model}

 In calculating the jet dynamics, we assume that the energy and baryon density 
within the ejecta do not have an angular dependence, and that the external 
medium is isotropic. We also assume that, at any time, the physical parameters
and bulk Lorentz factor $\Gamma$ are the same in the entire swept up external gas.
We include the effect of radiative energy losses on the jet dynamics.

 For the calculation of synchrotron emission, we assume a tangled magnetic
field and that the electrons accelerated at shock have a power-law distribution.
The electron distribution resulting from the continuous injection at shock and 
adiabatic+radiative cooling is approximated by broken power-law, with a break at 
the minimum random Lorentz factor of the freshly injected electrons and another 
one at the cooling electron Lorentz factor. In calculating the received flux, 
the swept up gas is approximated as a surface, \ie the thickness of the emitting 
shell is ignored. The effect of this surface curvature on the photon arrival time 
and energy is taken into account.

\subsection{Dynamics}
\label{dynamics}

 The interaction between the relativistic ejecta that generated the GRB with
the external gas continuously decelerates the jet and heats the newly swept
up gas. Assuming that the heated gas has a uniform temperature, equal to that
of the freshly shocked fluid, the total energy of the GRB remnant is:
\beq
 E(r) = m(r)(\Gamma^2-1) + m_0(\Gamma-1) \;,
\label{Gamma}
\eeq
where $m_0$ is the mass of the ejecta, whose Lorentz factor and energy at the end
of the GRB phase are $\Gamma_0$ and $E_0 = (\Gamma_0-1) m_0 c^2$, respectively. 

 In the above equation, $m$ is the mass of the swept up external gas, given by
\beq
 \dd m(r) = \Omega(r) \rho(r) r^2 \,\dd r \;,
\label{mjet}
\eeq
where $\rho(r) \propto r^{-s}$ is the external medium density ($s=0$ for 
homogeneous gas, $s=2$ for a wind ejected at constant speed before the release 
of the ultra-relativistic ejecta), and $\Omega(r)$ is solid angle of the jet.

 The jet half opening angle increases due to the lateral spreading of the jet 
at the local sound speed $c_s$:
\beq
 r \,\dd \theta = c_s \,\dd t' = \Gamma^{-1} c_s \,\dd t_{lab} \;,
\label{theta}
\eeq
$t'$ and $t_{lab}$ being the time measured in the ejecta comoving and laboratory 
frames, respectively. The speed of sound is (see Huang \etal 2000)
\beq
 c_s^2 = \frac{\adindex (\adindex-1) u'} {\rho' + \adindex u'}\, c^2 =
      \frac{\adindex (\adindex-1) (\Gamma-1)}{1+ \adindex(\Gamma-1)}\, c^2 \;,
\label{cs} 
\eeq
where $u'=(\Gamma-1) \rho'$ and $\rho'$ are the comoving internal energy and 
rest mass densities, respectively, and $\adindex$ is the adiabatic index. 
In the relativistic limit $\adindex = 4/3$ and $c_s = c/\sqrt{3}$, while for 
non-relativistic speeds $\adindex = 5/3$ and $c_s = (\sqrt{5}/3) v$, where
$v$ is the radial expansion speed. We relate the adiabatic index with $\Gamma$ 
through a simple formula, which has the above asymptotic limits. 

 The energy losses through synchrotron and inverse Compton emissions is given by
\begin{equation}
 \frac{\dd E}{\dd t_{lab}} = - \frac{\sigma_e c}{6\pi} \frac{m(r)}{m_p} B^2 (Y+1) 
              \int_{\gamma_m}^{\gamma_M} \gamma^2 \,\dd {\cal N}(\gamma)  \;,
\label{dEdr}
\end{equation}
where $B$ is the magnetic field intensity, $Y$ the Compton parameter,
${\cal N}$ is the normalized electron distribution (\S\ref{spectrum}), 
and $\gamma$ is the electron random Lorentz factor.

 Equations (\ref{Gamma}), (\ref{mjet}), (\ref{theta}) and (\ref{dEdr}) are solved
numerically, subject to the boundary conditions: $\Gamma(0)=\Gamma_0$, $m(0)=0$, 
$\theta(0)=\theta_0$, and $E(0)=E_0$.

\subsection{Electron Distribution and Spectral Breaks}
\label{spectrum}

The magnetic field intensity is parameterized relative to its equipartition value
\beq
 B^2 = 8\pi\, \rho' c^2 (\Gamma-1) \epsmag =
    32\pi\, \epsmag \, \rho(r) c^2 (\Gamma-1) \frac{\adindex \Gamma + 1}{\adindex -1} \;,
\label{Bmag}
\eeq
where $\rho'$ is the comoving frame rest-mass density.

 The distribution of the electrons accelerated by the forward shock and injected
in down-stream is assumed to be a power-law of index $p$
\beq
  {\cal N}_i (\gamma) \propto \gamma^{-p} \;, \quad \gamma_i < \gamma < \gamma_M \;,
\label{Ninj}
\eeq
where $\gamma_i$ is the minimum, injected electron Lorentz factor, parameterized 
relative to its value at equipartition,
\beq
 \gamma_i = \epsel \frac{m_p}{m_e} (\Gamma - 1) \;,
\label{gminj}
\eeq
and $\gamma_M$ is an upper limit, determined by the conditions that the acceleration
timescale of such electrons does not exceed the radiative losses timescale, and that
the total energy in the injected electrons does not exceed a certain fraction 
$\epsmax$ of the available internal energy. The former condition leads to
\beq
 \gamma_M^{(1)} = \left[ \frac{3\,e}{n_g \sigma_e} \frac{1}{B(Y+1)} \right]^{1/2} \;,
\label{gmmax1}
\eeq
where $n_g$ is the ratio of the acceleration timescale to the gyration time.
The latter condition can be written as 
\beq
 m_e \int_{\gamma_i}^{\gamma_M^{(2)}} \gamma \, \dd {\cal N}_i (\gamma) =
         \epsmax\, m_p (\Gamma-1) \;, 
\label{gmmax2}
\eeq
where ${\cal N}_i$ is normalized (to unity), and $m_e$ and $m_p$ are the electron 
and proton mass, respectively. Equation (\ref{gmmax2}) leads to an algebraic 
equation which can be solved numerically. The upper limit $\gamma_M$ is the minimum 
between $\gamma_M^{(1)}$ and $\gamma_M^{(2)}$ above. Unless $n_g$ is larger than about
$10^3$, the upper limit given in equation (\ref{gmmax1}) is sufficiently high that the 
synchrotron emission from $\gamma_M^{(1)}$-electrons is above the soft $X$-ray domain. 
However if $p \siml 2.5$ and $\epsmax$ is not much larger than $\epsel$, the upper limit 
given by equation (\ref{gmmax2}) may be sufficiently low to yield a break of the 
afterglow emission at the frequencies of interest ($X$-rays and even optical). 
For numerics we shall use $n_g=10$ and $\epsmax = 0.5$, the latter corresponding to
equipartition between electron and protons. 
 
 The distribution of cooled electrons is a power-law of index 2 if the electrons
are cooling faster than the dynamical timescale, and a power-law steeper by unity 
than the injected distribution in the opposite case (Sari, Piran \& Narayan 1998).
Therefore the electron distribution resulting from injection at shock and radiative 
cooling is 
\beq
 {\cal N}(\gamma) \propto  \left\{ \begin{array}{ll}
           \gamma^{-2}     & \gamma_c < \gamma < \gamma_i \\
           \gamma^{-(p+1)} & \gamma_i < \gamma < \gamma_M \end{array} \right.  \;,
\label{Nfast}
\eeq
for fast cooling electrons ($\gamma_c < \gamma_i$), and 
\beq
 {\cal N}(\gamma) \propto  \left\{ \begin{array}{ll}
           \gamma^{-p}     & \gamma_i < \gamma < \gamma_c \\
           \gamma^{-(p+1)} & \gamma_c < \gamma < \gamma_M \end{array} \right.  \;,
\label{Nslow}
\eeq
for slow cooling electrons ($\gamma_i < \gamma_c$). In equations (\ref{Nfast})
and (\ref{Nslow}), $\gamma_c$ is the cooling electron Lorentz factor, defined
by the equality of its radiative cooling timescale with the dynamical timescale:
\beq
 \gamma_c = 6\pi \frac{m_e c}{\sigma_e} \frac{1}{t' B^2 (Y+1)} \;.
\label{gmcool}
\end{equation}

The Compton parameter $Y$ is given by
\beq
 Y = \frac{4}{3} \tau_e \int_{\gamma_m}^{\gamma_M} \gamma^2 \dd {\cal N}(\gamma) \;,
\label{Ypar}
\eeq
where $\gamma_m = \min (\gamma_i, \gamma_c)$ and $\tau_e$ is the optical thickness
to electron scattering:
\beq
 \tau_e=\frac{\sigma_e}{m_p} \frac{m(r)}{\Omega(r) r^2} \;.
\label{tauel}
\eeq
The Klein-Nishina effect reduces the inverse Compton losses above an electron Lorentz
factor $\gamma_{KN}$ approximated as the geometric mean of $i)$ the electron Lorentz 
factor for which scattering of the synchrotron photons emitted by such an electron
occurs at the Klein-Nishina limit and of $ii)$ the electron Lorentz factor for which 
scattering of the synchrotron photons emitted by $\gamma_m$-electrons is at the same 
limit. The comoving frame synchrotron characteristic frequency for an electron of 
Lorentz factor $\gamma$ is 
\beq
 \nu' (\gamma) = \frac{3}{16} \frac{e}{m_e c}\, B \gamma^2 \;.
\label{nu}
\eeq
We take into account the Klein-Nishina reduction by calculating the integral in equation 
(\ref{Ypar}) up to $\gamma_{KN}$ if $\gamma_{KN} < \gamma_M$, and by switching off the
inverse Compton losses above $\gamma_{KN}$ in the integral given in equation (\ref{dEdr}).

The synchrotron self-absorption frequency in the fluid rest frame is at $\nu'_a =
\nu'(\gamma_a)$ with $\gamma_a$ given by (see Panaitescu \& Kumar 2000)
\beq
 \gamma_a = \left( \frac{5\,e}{\sigma_e} \frac{\tau_e}{B} \right)^{3/10} \gamma_m^{-1/2} \;.
\label{gmabs}
\eeq
This equation is valid only if $\gamma_a < \gamma_m$.

\subsection{Received Flux}
\label{observer}

 The synchrotron spectrum is approximated as piece-wise power-law (see Sari
\etal 1998) with breaks at the injection, cooling, and absorption breaks given 
by equations (\ref{nu}), (\ref{gminj}), (\ref{gmcool}), and (\ref{gmabs}). 

 To calculate the afterglow flux seen by the observer, we consider that the
emitting shell is infinitely thin and that the observer is located on the jet
axis. Consider an annular region of area $\delta A = 2\pi\, r^2 \delta \mu$,
with $\mu = \cos \omega$, where $\omega$ is the polar angle, measured relative 
to the jet axis. The energy emitted in the comoving frame per unit time and 
frequency is $\delta L'_{\nu'} = P'_{\nu'} \Sigma_e \delta A$, where $P'_{\nu'}$ 
is the radiative comoving power per electron and $\Sigma_e$ is the electron 
surface density. The infinitesimal comoving energy emitted per solid angle 
$(1/4\pi) \delta L'_{\nu'} \dd \nu' \dd t'$ is relativistically beamed toward 
the observer by a factor ${\cal D}^2$ and boosted in frequency by a factor 
${\cal D}$, where ${\cal D} = [\Gamma (1-\beta \mu)]^{-1}$, with $\beta = v/c$. 
Therefore the infinitesimal flux $\dd F_\nu$ received by the observer at 
frequency $\nu = {\cal D} \nu'$, during $\Delta t = c^{-1} r \delta \mu$ satisfies
\beq
 \dd F_\nu\, \dd \nu\, \delta t = \frac{1+z}{4\pi\,d_L^2} {\cal D}^3 P'_{\nu'}
                                 \Sigma_e\, \dd t'\, \dd \nu'\, \delta A \;,
\label{Fnu1}
\eeq
where $P'_{\nu'}$ is the sum of synchrotron and inverse Compton emissions, $z$ is 
the afterglow redshift, and $d_L$ is the luminosity distance. We assume a Universe 
with $H_0 = 65\; {\rm km \, s^{-1} Mpc^{-1}}$, $q_0 = 0.1$, and $\Lambda =0$.

 The flux received by the observer at time $t$ is that given by equation (\ref{Fnu1}), 
integrated over the entire evolution of the source. 
Using $\dd t' = \dd t_{lab}/ \Gamma = \dd r / (\beta c \Gamma)$ and relating the 
electron surface density to the jet mass and area, $\Sigma_e = m(r)/(m_p \Omega r^2)$, 
equation (\ref{Fnu1}) leads to 
\beq
  F_\nu (t) = \frac{1+z}{2\,m_p d_L^2} \int 
              \frac{\dd r}{\beta\Gamma^3 (1-\beta \mu)^2}  
              \frac{P'_{\nu'}(r) m(r)}{r \Omega(r)} \;, 
\label{Fnu}       
\eeq
with $\mu$ given by the condition that light emitted from location $(r, \mu)$ 
arrives at observer at time $t = t_{lab} - (r/c) \mu$. Thus equation (\ref{Fnu}) 
takes into account the spread in the photon arrival time due to the spherical 
curvature of the jet surface. In all our calculations it is assumed that the 
observer is on the jet axis.

\section{Analytical Considerations}
\label{boldtitle}

 So far there are five afterglows (990123, 990510, 991216, 000301c, 000926) for which 
a break in the optical emission has been identified. In all these cases the break is 
seen at or after $t=1$ day.  Within the framework of uniform ejecta interacting with 
isotropic media, there are two possible causes for such a break: the passage of a break 
frequency (injection, $\nu_i$, cooling, $\nu_c$, or that due the upper cut-off of the 
electron distribution, $\nu_M$), or the edge of the jet becoming visible to the observer
(plus the changing dynamics due to the lateral spreading of the jet). 

 One can show that, within a factor of order unity, the break frequency $\nu_i$ 
is the same for both types of external medium:
\beq
 \nu_i \sim 10^{13}\; (z+1)^{1/2} {\cal E}_{0,54}^{1/2} \varepsilon_{e,-1}^2 
                      \varepsilon_{B,-2}^{1/2} t_d^{-3/2}\; {\rm Hz}\;,
\label{nui}
\eeq
where ${\cal E}_0 = 4\pi (E_0/\Omega_0)$ is the jet isotropic equivalent energy,
$t_d$ is observer time in days, and the usual notation $A_n=10^{-n} A$ was used. 
Equation (\ref{nui}) shows that, unless the isotropic equivalent energy 
exceeds $10^{56}$ erg and the magnetic field is close to equipartition 
($\epsel$ cannot be much higher than 0.1, as the fractional energy in electrons 
must be below unity), $\nu_i$ is below the optical range at $t \simg 1$ day. 
Therefore it is very unlikely that the optical light-curve breaks are due to the passage 
of $\nu_i$ through the observational band. Moreover, if this were the case, then, 
at times before the light-curve break, the temporal index $\alpha$ of the light-curve 
decay, $F_\nu (t) \propto t^{-\alpha}$, would be at most $1/4$, which is much smaller 
than the observed $\alpha$'s.

\subsection {Passage of the Cooling Break}
\label{cooling}

We consider here the afterglow emission at early times, when the effects due to 
collimation of ejecta are negligible, but sufficiently large that $\nu_i < \nu$. 
In this case, the afterglow light-curves for slow cooling electrons ($\nu_i < \nu_c$) 
are given by (see Panaitescu \& Kumar 2000) 
\beq
 F_{\nu>\nu_i} \propto \left\{ \begin{array}{ll}
           t^{-(3p-3)/4} &  \nu < \nu_c \\
           t^{-(3p-2)/4} &  \nu_c < \nu,\; Y < 1 \\
           t^{-(3p/4)+1/(4-p)} & \nu_c < \nu,\; Y > 1,\; 2 < p < 3 \\
           t^{-(3p/4)+1} & \nu_c < \nu,\; Y > 1,\; 3 < p \\
          \end{array} \right.
\label{Fs0}
\eeq
for a homogeneous external medium ($s=0$), where the last two rows represent the
case when the electron cooling is dominated by inverse Compton scatterings. 
For a wind-like medium ($s=2$) and slow cooling electrons
\beq
 F_{\nu>\nu_i} \propto \left\{ \begin{array}{ll}
           t^{-(3p-1)/4} & \nu < \nu_c \\
           t^{-(3p-2)/4} &  \nu_c < \nu,\; Y < 1 \\
           t^{-(3p/4)+p/(8-2p)} & \nu_c < \nu,\; Y > 1,\; 2 < p < 3 \\
           t^{-(3p/4)+3/2} & \nu_c < \nu,\; Y > 1,\; 3 < p \\
          \end{array} \right.
\label{Fs2}
\eeq
The second row of equations (\ref{Fs0}) and (\ref{Fs2}) also gives the light-curve
for $\nu_i < \nu$ and fast cooling electrons.

 The temporal evolution of the cooling break frequency $\nu_c$ for $s=0$ and slow 
cooling electrons is given by
\beq
 \nu_c \propto  \left\{ \begin{array}{ll}
        t^{-1/2} & Y < 1 \\
        t^{-(8-3p)/(8-2p)} & Y > 1 \;,\; 2 < p < 3 \\
        t^{1/2} & Y > 1  \;,\; 3 < p \\
        \end{array} \right. \;.
\label{nucs0}
\eeq
For $s=2$ and slow cooling electrons
\beq
 \nu_c \propto  \left\{ \begin{array}{ll}
        t^{1/2} & Y < 1 \\
        t^{(3p-4)/(8-2p)} & Y > 1 \;,\; 2 < p < 3 \\
        t^{5/2} & Y > 1  \;,\; 3 < p \\
        \end{array} \right. \;.
\label{nucs2}
\eeq
The first row also gives the evolution of $\nu_c$ for fast cooling electrons, 
in which case $Y$ is time-independent. 

 For a homogeneous medium, equation (\ref{nucs0}) shows that $\nu_c$ increases in time 
if the electron cooling is dominated by inverse Compton and if $p > 8/3$. From equation 
(\ref{Fs0}), the passage of $\nu_c$ through the observational band changes the 
light-curve decay index by
\beq
 (\Delta \alpha)_c =  \left\{ \begin{array}{ll}
         1/4 &  Y < 1 \\
        (8-3p)/(16-4p) & Y > 1 \;,\; 2 < p < 8/3 \\
        (3p-8)/(16-4p) & Y > 1 \;,\; 8/3 < p < 3 \\
         1/4 &  Y > 1 \;,\; 3 < p  \\ 
             \end{array} \right. \;.
\label{Das0}
\eeq
Note $(\Delta \alpha)_c > 0$, \ie the passage of $\nu_c$ always steepens the 
light-curve decay, even if $\nu_c$ increases in time, and that $(\Delta \alpha)_c < 1/4$. 
In the case where $\nu_c$ is above optical and below $X$-ray, the temporal indices of
the $X$-ray and optical light-curves differ by $\alpha_x-\alpha_o=(\Delta \alpha)_c > 0$ 
if $\nu_c$ decreases in time, and by $\alpha_x-\alpha_o = -(\Delta \alpha)_c < 0$ if 
$\nu_c$ increases in time. Thus, for $s=0$, the $X$-ray emission decays faster 
than the optical one if $Y < 1$ or $Y > 1$ and $p < 8/3$.

 For a wind-like medium equation (\ref{nucs2}) shows that $\nu_c$ always increases in time.
From equation (\ref{Fs2}), the passage of $\nu_c$ yields
\beq
 (\Delta \alpha)_c =  \left\{ \begin{array}{ll}
         1/4 &  Y < 1 \\
        (3p-4)/(16-4p) & Y > 1 \;,\; 2 < p < 3 \\
         5/4 &  Y > 1 \;,\; 3 < p  \\
             \end{array} \right. \;.
\label{Das2}
\eeq
Note that $1/4 < (\Delta \alpha)_c < 5/4$. For $\nu_o < \nu_c < \nu_x$, the $X$-ray 
and optical indices differ by $\alpha_x - \alpha_o = -(\Delta \alpha)_c < 0$. 
Hence, for $s=2$, the $X$-ray emission always decays slower than the optical one.

\subsection{Collimation of Ejecta}
\label{collimation}

 If the ejecta is collimated, the decay of the afterglow emission steepens around the time 
$t_j$ when $\Gamma \theta = 1$, due to the altered jet dynamics and that the observer sees 
the edge of the jet. For $s=0$
\beq
 t_j = 1.2\;(z+1) \left( {\cal E}_{0,54} \theta_{0,-1}^8 n_0^{-1} \right)^{1/3} \;{\rm day}\;.
\label{tjet}
\eeq
The coefficient above was determined numerically from the arrival time of the photons moving
toward the observer along the jet axis. Photons emitted from other regions on the jet surface
arrive later by a factor up to $\sim 4$.
 
 Around $t_j$ the jet dynamics changes from a quasi-spherical expansion with $\Gamma \propto 
r^{-3/2} \propto t^{-3/8}$ for $s=0$ ($\Gamma \propto r^{-1/2} \propto t^{-1/4}$ for $s=2$) 
to a sideways expansion characterized by $\Gamma \propto e^{-kr} \propto t^{-1/2}$ (Rhoads 
1999). During the lateral spreading phase ($t > t_j$) the cooling frequency evolution is 
\beq
 \nu_c \propto  \left\{ \begin{array}{ll}
        t^0 & Y < 1 \\
        t^{2(p-2)/(4-p)} & Y > 1 \;,\; 2 < p < 3 \\
        t^2 & Y > 1  \;,\; 3 < p \\
        \end{array} \right. \;,
\label{nucjet}
\eeq
assuming slow cooling electrons ($\nu_i < \nu_c$). Then it can be shown that, at $t > t_j$, 
the light-curve is given by
\beq 
 F_{\nu > \nu_i}  \propto  \left\{ \begin{array}{lll}
        t^{-p}              & {\rm any}\; \nu \;, & Y < 1 \\
        t^{-p +(p-2)/(4-p)} & \nu_c < \nu \;,     & Y > 1 \;,\; 2 < p < 3 \\
        t^{-(p-1)}          &  \nu_c < \nu \;,    & Y > 1  \;,\; 3 < p \\
        \end{array} \right. \;.
\label{Fjet}
\eeq
Evidently, as the source slows down, the $Y$ parameter eventually falls below unity
and the last two cases given in equation (\ref{Fjet}) approach asymptotically $F_\nu \propto
t^{-p}$. The results given in equations (\ref{nucjet}) and (\ref{Fjet}) ignore multiplying 
terms that are powers of the jet radius $r$, which increases logarithmically with the observer 
time. With the same approximation, they also hold for a jet interacting with a pre-ejected 
wind. Furthermore, these results are accurate only at times when the afterglow is very 
relativistic.  From numerical calculations we found that, for the first case given in equation 
(\ref{Fjet}), the decay index $\alpha$ is approximated by $p$ with an error less than 10\% if 
$\Gamma \simg 10$. 

 Using equations (\ref{Fs0}), (\ref{Fs2}), and (\ref{Fjet}) it can be shown that, for
$\nu > \nu_i$, the magnitude of the break due to collimation of ejecta is
\beq
 (\Delta \alpha)_j =  \left\{ \begin{array}{ll}
         (p+3)/4 &  \nu < \nu_c \\
         (p+2)/4 &  \nu_c < \nu, \; Y < 1 \\
         p/4+1/(4-p) & \nu_c < \nu,\; Y > 1,\; 2 < p < 3 \\
         (p+4)/4 &  \nu_c < \nu,\; Y > 1,\; 3 < p \\
        \end{array} \right. \;.
\label{das0}
\eeq
for $s=0$ and
\beq
 (\Delta \alpha)_j =  \left\{ \begin{array}{ll}
         (p+1)/4 &  \nu < \nu_c \\
         (p+2)/4 &  \nu_c < \nu, \; Y < 1 \\
         p/4+p/(8-2p) & \nu_c < \nu,\; Y > 1,\; 2 < p < 3 \\
         (p+6)/4 &  \nu_c < \nu,\; Y > 1,\; 3 < p \\
        \end{array} \right. \;.
\label{das2}
\eeq
for $s=2$. The finite opening of the ejecta yields $\Delta \alpha = 3/4$ for $s=0$
and $\Delta \alpha = 1/2$ for $s=2$ (Panaitescu, \Meszaros \& Rees 1998) when the 
jet edge becomes visible, the remainder of the steepening being due to the sideways 
expansion of the jet (Kumar \& Panaitescu 2000). Equations (\ref{das0}) and (\ref{das2}) 
show that, if $p > 2$, $(\Delta \alpha)_j > 1$ for $s=0$ and $(\Delta \alpha)_j > 3/4$
for $s=2$ .

\subsection{What Can We Infer from the $X$-ray \\ and Optical Decay Indices ?}
\label{comparison}

 The most important difference between a break due to passage of $\nu_c$ and one 
due to collimation of ejecta is the chromaticity of the former and the achromaticity 
of the latter. This would be the best criterion to distinguish between
them if optical and $X$-ray observations spanning the same 1--2 decades in time,
around the time when the break is seen, are available. 

The analytical results presented in \S\ref{cooling} and \S\ref{collimation} allow us 
to draw some conclusions even when the existence of simultaneous $X$-ray and optical 
light-curve breaks cannot be clearly established. Equations (\ref{Das0}), (\ref{Das2}), 
(\ref{das0}), and (\ref{das2}) show that optical break magnitudes $\Delta \alpha < 3/4$ 
can be produced {\sl only} by the passage of the cooling break, while $\Delta \alpha > 5/4$ 
can be due {\sl only} to collimation of ejecta. The caveat of this criterion is that, 
as shown by Kumar \& Panaitescu (2000), for collimated ejecta, the completion of most 
of $(\Delta \alpha)_j$ is spread over at least one decade in observer time for $s=0$ 
and over at least two decades for $s=2$. Therefore observations spanning a shorter time 
range will underestimate the true magnitude of the jet edge break, particularly in the
case when the observer is not located close to the jet axis. 

 The results presented in \S\ref{cooling} and \S\ref{collimation} also lead to the 
following criteria for determining the location of $\nu_c$ relative to the optical
and $X$-ray domains, and for identifying the type of external medium, from the optical
and $X$-ray decay at the {\sl same} time: 
\begin{enumerate}
\item \vspace*{-2mm}
   if $\alpha_o = \alpha_x$, then either $\nu_c$ is not between the optical and 
   $X$-ray domains, or it is in this range but is quasi-constant in time. The latter
   may occur either in collimated ejecta during the sideways expansion phase, or 
   before jet edge effects are important, in $s=0$ models, if the inverse Compton 
   cooling is the dominant process and if $p \simeq 8/3$ (see \eq [\ref{nucs0}]). 
\item \vspace*{-2mm}
   if $\alpha_x > \alpha_o$ (\ie the $X$-ray decay is faster than in the optical), 
   then $\nu_o < \nu_c < \nu_x$ and the external medium is homogeneous. Furthermore
   either the electron cooling is synchrotron dominated or, if Compton-dominated, 
   $p < 8/3$. 
\item \vspace*{-2mm}
   if $\alpha_x < \alpha_o$, then $\nu_o < \nu_c < \nu_x$ and either the external
   medium is a wind-like or it is homogeneous and the electron cooling is inverse 
   Compton--dominated ($Y > 1$), with $p > 8/3$. 
\end{enumerate}
\vspace*{-2mm}
Using the observed spectral slope in the optical range and the relative intensity
of the $X$-ray and optical emission, one can eliminate some of the above cases,
and further reduce the number of potentially good models for a given afterglow.

\section{Individual Afterglows}
\label{aglows}

 The modeling of GRB afterglow light-curves is carried out by solving numerically 
equations (\ref{Gamma}), (\ref{mjet}), (\ref{theta}), and (\ref{dEdr}), to determine 
the dynamics of the afterglow, and equation (\ref{Fnu}) to calculate the observed flux. 
The six unknown parameters ${\cal E}_0$, $\theta_0$, $n$, $p$, $\epsel$, and $\epsmag$ 
are determined by minimizing the $\chi^2$ between observed and model fluxes at the 
frequencies where most of the data is available. 
The jet initial Lorentz factor $\Gamma_0$ does not affect the afterglow emission if 
its evolution is quasi-adiabatic, because it drops out from the calculation of the 
jet Lorentz factor $\Gamma$ as a function of observer time. When radiative losses 
are taken into account, $\Gamma_0$ may have an effect on the afterglow emission, as 
the magnitude of these losses depends on $\Gamma_0$. However, the effect is weak and 
cannot be used to set significant constraints on $\Gamma_0$. For simplicity, we keep
a fixed $\Gamma_0 = 500 $ in all numerical calculations.

 In this work we restrict our attention to four afterglows for which radio, optical, 
$X$-ray light-curves and redshifts are available: GRB 980703, GRB 990123, GRB 990510, 
and GRB 991216, leaving out GRB 970508, whose optical light-curve cannot be entirely
explained within the framework of our model, as it exhibited a brightening after 1 day,
indicating a possible delayed energy injection, or fluctuations of the energy release
parameters $\epsel$, $\epsmag$ or of the external density $n$. 

 The near-infrared (NIR) and optical magnitudes are converted to fluxes using the 
characteristics of photometric bands published by Campins, Rieke, \& Lebofsky (1985), 
and Fukugita, Shimasaku \& Ichikawa (1995). We assumed a 5\% uncertainty in the 
magnitude--flux conversion. The NIR and optical fluxes are corrected for dust extinction 
using the interstellar reddening curves of Schild (1977), Cardelli, Clayton, \& Mathis 
(1989), and Mathis (1990). A 10\% uncertainty is assumed for the Galactic extinction.

\subsection{GRB 980703}
\label{980703}
 
 The emission of the afterglow of GRB 980703 is dominated by the host galaxy
at only few days after the main event. No break has been detected in the optical
within this time interval, which means that we can set only a lower limit on the 
jet opening angle, by requiring $t_j$ to be sufficiently large. 

 The decay of the optical emission is characterized by a power-law index $\alpha_o = 
1.17 \pm 0.25$ (Bloom \etal 1998), or $\alpha_o = 1.39 \pm 0.30$ (Castro-Tirado 
\etal 1999a), or $\alpha_o = 1.63 \pm 0.12$ (Vreeswijk \etal 1999), thus the average 
index is $\overline{\alpha_o} = 1.53 \pm 0.10$. The slope of the optical spectrum is 
$\beta_o = 2.71 \pm 0.12$ (Vreeswijk \etal 1999) at $t=1.2$ day, and an $X$-ray 
decay index $\alpha_x = 1.33 \pm 0.25$ was found by Galama \etal (1998). Thus the 
observations give $\alpha_x-\overline{\alpha_o} = -0.20 \pm 0.27$. In view of the 
analytical considerations given in \S\ref{cooling}, the above $\alpha_x - 
\overline{\alpha_o}$ marginally rules out only a model with $s=0$ and $\nu_c < \nu_x$. 

 For $s=0$ and $\nu_o < \nu_c$, the optical decay index is $\alpha_o = (3p-3)/4$, thus
observations require that $p=3.15 \pm 0.16$, which implies $\beta_o = (p-1)/2 = 1.07 \pm 
0.08$. Such a spectrum 
is much harder than observed, therefore consistency between observations and the 
fireball model can be achieved only if there is a substantial extinction. Given that 
the Galactic extinction toward this afterglow is $E(B-V)=0.061$ (Bloom \etal 1998), 
it follows that most this extinction is due to the host galaxy. Vreeswijk \etal (1999) 
have shown that the synchrotron power-law spectrum becomes consistent with the observed 
one for an intrinsic extinction of $A_V = 1.45 \pm 0.13$. At the same time the dereddened 
optical fluxes lead to $\beta_{ox} = 1.06 \pm 0.04$, which is consistent with the 
dereddened $\beta_o = (p-1)/2$, implying that the cooling break is above $X$-rays. 

 For $s=2$ and assuming $\nu_o < \nu_c < \nu_x$, leads to $\alpha_o = (3p-1)/4$,
therefore the observed $\alpha_o$ requires $p=2.48 \pm 0.16$ and $\beta_o = (p-1)/2 =
0.74 \pm 0.08$, which is again much harder than observed. Using the approximations 
found by Cardelli, Clayton, \& Mathis (1989) for the UV interstellar reddening,
one can show that dust extinction in the host galaxy steepens the 
synchrotron spectrum by $\Delta \beta_o \simeq 1.13 A_V$ at the frequency which 
is red-shifted in the observer $V$-band. Therefore the synchrotron and observed 
spectra are consistent if $A_V = 1.74 \pm 0.13$, which is larger than for the $s=0$ model
because a wind-like medium yields softer spectra for the same decay index $\alpha$. 
The corresponding dereddened optical emission has $\beta_{ox} = 1.11 \pm 0.04$, which 
falls between the limiting values, $(p-1)/2$ and $p/2$, allowed by the synchrotron model.

 Thus the afterglow of GRB 980703 may be explained by models with homogeneous external
media, $\nu_o < \nu_x < \nu_c$, $p \sim 3.1$, and an intrinsic extinction $A_V = 1.45 
\pm 0.13$, or by wind models with $\nu_o < \nu_c < \nu_x$, $p \sim 2.5$, and $A_V = 1.74 
\pm 0.13$. Figure 1 shows an $s=0$ model yielding an acceptable fit to the data.

\subsection{GRB 990123}
\label{990123}
 
 After the subtraction of the host galaxy, the $r$-band light-curve of the afterglow of 
GRB 990123 had a break around $t_b=2$ day (Kulkarni \etal 1999a), with the asymptotic 
logarithmic slope changing by $\Delta \alpha_o \simeq 0.55 \pm 0.07$ from that at early 
times, $\alpha_{o,1} = 1.10 \pm 0.03$, to $\alpha_{o,2}=1.65\pm 0.06$ after $t_b$.
The same break magnitude is implied by the power-law indices found by 
Castro-Tirado \etal (1999b). The reported slopes of the optical spectrum are $\beta_o 
= 0.75 \pm 0.23$ at $t=1.2$ day (Galama \etal 1999), and $\beta_o = 0.8 \pm 0.1$ at
$t=1$ day (Kulkarni \etal 1999a). The optical-to-$X$-ray spectral slope $\beta_{ox}$ 
reported by Galama \etal (1999) is $\beta_{ox}= 0.67 \pm 0.02$ at $t=1.2$ day. Therefore 
$\beta_{ox} - \beta_o = -0.1 \pm 0.1$, indicating that the cooling break is not between 
optical and $X$-ray frequencies at $t \siml t_b$. The first BeppoSAX measurement (Heise 
\etal 1999) and the ASCA data (Murakami \etal 1999) give an $X$-ray decay index 
$\alpha_x=1.17 \pm 0.10$ for $0.2\,{\rm d} < t < 2\,{\rm d}$, therefore $\alpha_x \simeq 
\alpha_{o,1}$, consistent with both $\nu_c < \nu_o$ and $\nu_x < \nu_c$.

 If the break in the optical emission were due to the passage of the $\nu_c$, then
$\Delta \alpha_o > 1/4$ requires $s=2$ and an increasing $\nu_c$ (see \S\ref{cooling}).
Equation (\ref{Das2}) and the observed $\Delta \alpha_o$ give $p = 2.46 \pm 0.08$.
Then at $t < t_b$, when $\nu_c < \nu_o$, the optical-to-$X$-ray slope should be 
$\beta_{ox} = p/2 = 1.23 \pm 0.04$, clearly inconsistent with the observations. 
This shows that the optical break seen in this afterglow is not caused by the passage 
of $\nu_c$. 

 A cooling frequency below optical cannot be accommodated by a jet model either, 
as the observed $\beta_{ox}$ would imply $p = 2\beta_{ox} = 1.34 \pm 0.04$, leading 
to a decay index $\alpha_{o,1} = (3p-2)/4 = 0.51 \pm 0.03$ (irrespective of the type 
of external medium) inconsistent with the observed value. Therefore the cooling frequency 
must be above $X$-ray, implying $p = 2\beta_{ox} + 1 = 2.34 \pm 0.04$. Then $s=0$ leads 
to $\alpha_{o,1} = (3p-3)/4 = 1.00 \pm 0.03$, while $s=2$ yields $\alpha_{o,1} = (3p-1)/4 
= 1.50 \pm 0.03$. The latter case is clearly inconsistent with the observations.

 From this analysis one can conclude that a successful model for the afterglow 
of GRB 990123 is that of a jet interacting with a homogeneous external medium, with
parameters such that $\nu_o < \nu_x < \nu_c$ at $t < t_b$, and $p \sim 2.3$ . 
This value of $p$ implies an optical decay index $\alpha_{o,2} \sim p$ substantially 
larger than found by Kulkarni \etal (1999a). However, the observations made after 
$t_j \sim 2$ day do not span a sufficiently long time and may underestimate the 
asymptotic $\alpha_{o,2}$.

 The radio and optical data at $t=1.2$ day give a radio-to-optical slope $\beta_{ro} = 
0.27 \pm 0.04$. If the injection break were below radio frequencies ($\nu_r$) at this 
time, then $p = 2 \beta_{ro} + 1 = 1.54 \pm 0.08$, inconsistent with the value 
derived above from the optical and $X$-ray data.  Therefore $\nu_r < \nu_i$ at $t < t_j$. 
But in this case, as pointed out by Kulkarni \etal (1999a), the radio emission should 
rise until $\nu_i$ falls below $\nu_r$. For a jet, this rise stops around $t_j$ and is
followed by a constant emission until $\nu_i = \nu_r$, after which the radio flux should
decrease. However, the radio emission of 990123 exhibits a strong dimming around 2 days
(see Figure 2) which, as suggested by the argument above, cannot be explained by the
forward shock emission in a jet model (see the radio light-curve shown in Figure 2 with 
dashed line). We shall assume that the two earliest radio fluxes are produced by another 
radiation mechanism, for instance emission by cooled electrons that were accelerated by 
the reverse shock (Sari \& Piran 1999) or emission from less relativistic ejecta 
surrounding the jet, and use these fluxes only as upper limits for the forward shock 
emission. 

 Figure 2 shows a jet model for the emission of 990123. The $K$-band fluxes observed after 
10 days lie above the model prediction, being inconsistent with an achromatic break 
resulting from collimation of ejecta. If the optical flash of GRB 990123 was due to a 
reverse shock propagating in the ejecta (Sari \& Piran 1999), and if the peak of this flash, 
seen at $t \sim 50$ seconds, corresponds to the fireball deceleration timescale, then the 
isotropic equivalent energy and external density of the model in Figure 2 imply that 
the fireball initial Lorentz factor was $\Gamma_0 = 1400 \pm 700$. Alternatively, the 
optical flash may have been caused by internal shocks in an unstable wind (\Meszaros \& 
Rees 1999, Kumar \& Piran 2000), in which case $\Gamma_0$ is more uncertain.

\subsection{GRB 990510}
\label{990510}

 The optical afterglow of GRB 990510 exhibited a break around $t_b =1.5$ day, whose
reported magnitude is $\Delta \alpha_o = 1.80 \pm 0.20$ (Israel \etal 1999) in the 
$V$-band, $\Delta \alpha_o = 1.67 \pm 0.02$ (Stanek \etal 1999), to $\Delta \alpha_o = 
1.36 \pm 0.05$ (Harrison \etal 1999). The optical asymptotic decay indices found 
by Harrison \etal (1999) are $\alpha_{o,1} = 0.82 \pm 0.02$ and $\alpha_{o,2} = 2.18 
\pm 0.05$. The $X$-ray decay index was $\alpha_x = 1.42 \pm 0.07$ (Kuulkers \etal 1999) 
at $0.3\,{\rm d} < t < 2\,{\rm d}$, while the optical spectral slope was $\beta_o = 
0.61 \pm 0.12$ (Stanek \etal 1999) at $t=0.9$ day. The available data imply an
optical-to-$X$-ray slope $\beta_{ox} = 0.90 \pm 0.04$ at $t=0.72$ day.

 Even the smallest reported $\Delta \alpha_o$ is above $5/4$, therefore the break seen 
in the optical emission cannot be due to the passage of $\nu_c$ (see \S\ref{comparison}) 
and must have been caused by jet effects. From $\beta_{ox} - \beta_o = 0.29 
\pm 0.13$ one can infer that $\nu_c$ is between optical and $X$-rays. The same conclusion 
is suggested by $\alpha_{o,1} < \alpha_x$, however the $X$-ray observations were made 
at times close to $t_j$, thus the observed $X$-ray decay index may overestimate the 
asymptotic $\alpha_x$ at earlier times. 
 
 For a homogeneous medium $\alpha_{o,1} = (3p-3)/4$, therefore observations require that 
$p = 2.09 \pm 0.03$, thus the analytically expected values of $\alpha_{o,2} \sim p$ and 
$\beta_o= (p-1)/2 = 0.55 \pm 0.02$ are consistent with those observed. For a wind-like
medium $\alpha_{o,1} = (3p-1)/4$, requiring that $p = 1.43 \pm 0.03$, which leads to
an optical slope $\beta_o = (p-1)/2 = 0.22 \pm 0.02$ inconsistent even with the softest 
optical spectrum reported by Stanek \etal (1999): $\beta_o = 0.46 \pm 0.08$. Such a small
value of $p$ also implies $\beta_{ox} \leq p/2 = 0.72 \pm 0.02$, again inconsistent with 
the observations. Thus a wind-like medium is ruled out.

 Therefore the afterglow of GRB 990510 can be accommodated by a model with $s=0$, $\nu_o 
< \nu_c < \nu_x$ and $p \sim 2.1$. Note that the model shown in Figure 3 fits well the 
$X$-ray data, the light-curve steepening being very slow.

\subsection{GRB 991216}
\label{991216}

 The optical decay index of the afterglow of GRB 991216 was $\alpha_{o,1} = 1.22 \pm 
0.04$ (Halpern \etal 2000, Sagar \etal 2000) at $t < t_b = 2$ day. Halpern \etal
have shown that a broken power-law optical light-curve with $\alpha_{o,2} = 1.53 \pm 
0.05$ at $t > t_b$ plus the contribution $R = 24.8 \pm 0.1$ from a galaxy located at
$\sim 1$ arcsecond from the optical transient explains well the observations.
Sagar \etal (2000) find that two measurements made after $t_b$ are dimmer by $2\sigma$
than the power-law extrapolation from earlier fluxes. Therefore there is evidence
that the decay of the optical emission of the 991216 afterglow steepened by 
$\Delta \alpha_o = 0.31 \pm 0.06$

 The $X$-ray data span 1.3 decades in time before $t_b$ and have $\alpha_x=1.62 \pm 
0.07$ (Frail \etal 2000, Halpern \etal 2000), leading to $\alpha_x - \alpha_{o,1} = 
0.40 \pm 0.08$. The optical spectrum is puzzling, exhibiting a turn-over at the $J$-band 
at $t \simeq 1.5$ day (Frail \etal 2000, Halpern \etal 2000), although the $J$ and 
$K$-band measurements reported by Garnavich \etal (2000) restore a power-law spectrum 
of slope $\beta_o = 0.58 \pm 0.08$ at $t=1.7$ day. Garnavich \etal (2000) have pointed
out that, if extinction is overestimated by a factor 1.3--1.5 close to the Galactic plane
(Stanek \etal 1999), the dereddened optical spectrum could be softer: $\beta_o = 0.87 \pm 
0.08$. The optical-to-$X$-ray spectral slope is $\beta_{ox} = 0.80 \pm 0.10$ (Garnavich 
\etal 2000, Halpern \etal 2000) at $t=1.7$ day.

 The faster decay seen in the $X$-rays than in the optical before $t_b$ requires $\nu_o < 
\nu_c < \nu_x$ and a homogeneous external medium (see \S\ref{comparison}), in which case 
it is expected that $\alpha_x - \alpha_{o,1} \leq 1/4$. This difference is $2\sigma$ below 
the observed value. For $\nu_o < \nu_c$, the analytical optical decay index is $\alpha_{o,1} = 
(3p-3)/4$, thus observations imply $p=2.63 \pm 0.05$. This leads to $\beta_o = (p-1)/2 = 
0.82 \pm 0.03$, which is consistent with the softer dereddened suggested by Garnavich \etal
(2000). However, for such a spectrum, $\beta_{ox} - \beta_o = - 0.07 \pm 0.13$ does not 
support a cooling frequency between optical and $X$-rays. 

 If the optical light-curve break were due to the passage of $\nu_c$ then $\Delta \alpha_o = 
1/4$, consistent with the observations. Then at $t \sim t_b$, when $\nu_c \sim \nu_o$, the 
optical-to-$X$-ray slope should be $\beta_{ox} = p/2 = 1.32 \pm 0.03$, clearly inconsistent 
with the observations. Therefore the break in the optical emission of 991216 was not caused 
by the passage of $\nu_c$. Instead, the small magnitude optical break could be explained by
jet effects if the short time baseline of the observations after $t_j$ capture only a part 
of the full steepening. 

 However there are some major difficulties that a jet model with  $\nu_o < \nu_c < \nu_x$
encounters. The decay index of the radio emission $\alpha_r = 0.82 \pm 0.02$ (Frail \etal 2000) 
at $t \simg 1$ day requires the injection break $\nu_i$ to be below 10 GHz, which leads to a 
radio-to-optical spectral slope $\beta_{ro} = \beta_o = 0.6 \div 0.9$, while observations give 
$\beta_{ro} = 0.20 \pm 0.05$ at 2 days. Furthermore, the decay indices at radio and optical 
frequencies should be the same, yet observations show that $\alpha_{o,1} - \alpha_r = 0.40 \pm 
0.04$. The discrepancy is large enough to suggest that a model with the above features cannot 
accommodate the optical and all the radio data. A possible solution is to ``decouple"
$\alpha_r$ and $\alpha_{o,1}$ by assuming that the radio emission until several days has a 
different origin, as we assumed for the early radio emission of 990123. Then the quasi-flat 
behavior of the radio data at 8--50 days (Figure 4) is suggestive of a jet undergoing lateral 
spreading and with $\nu_i$ above radio until 50 days, as this is the only way of obtaining a 
constant flux if the external medium is homogeneous (the analytical prediction for $\nu < \nu_i$ 
and after the jet edge is seen, $F_\nu \propto t^{-1/3}$, is based on some approximations that 
are not sufficiently accurate; numerically we obtain a slightly different behavior $F_\nu \sim 
const$). 

 There is, however, a model which does not require another emission mechanism besides the 
forward shock or a two-component structure of the jet (Frail \etal 2000), and which can
accommodate the entire radio data. The almost constant flux exhibited by the radio emission 
at 1--3 days (Figure 4) can be explained by jet expanding laterally before 1 day, with the 
steepening of the radio emission at several days being due to the $\nu_i$-passage. Then 
the steepening of the optical decay at few days is not due to jet effects, as suggested by
Halpern \etal (2000), but to the passage of a spectral break. The cooling frequency $\nu_c$ 
does not offer a self-consistent picture: it should be above optical shortly before 2 days, 
as required by $\alpha_x > \alpha_{o,1}$ and, according to equation (\ref{nucjet}), it should 
increase in time because during the jet sideways expansion phase, thus it cannot cross the 
optical domain. 
 
  Within our afterglow modeling there is only one remaining possibility: the steepening of 
the optical emission is due to the passage of the $\nu_M$ frequency associated with the 
high energy break of the electron distribution (\eq [\ref{gmmax2}]). As the jet model yields 
$F_\nu \propto t^{-p}$ at all frequencies above $\nu_i$ (\eq [\ref{Fjet}]), the observed 
$\alpha_{o,1}$ implies a very hard electron distribution with $p \sim 1.2$. Then the observed 
$\beta_o = 0.58 \pm 0.08$ requires $\nu_c < \nu_o$ at 2 days, to obtain consistency with the 
analytical expectation $\beta_o = p/2 \sim 0.6$.
 
 Such a jet model with a hard electron distribution and cooling frequency below optical 
is shown in Figure 4. The observed optical emission steepens only mildly after 2 days, 
therefore the high energy break at $\gamma_M$ is not too strong. For simplicity we have
approximated this break as a softening of the electron distribution from $\gamma^{-p}$
to $\gamma^{-(p+\delta p)}$ at $\gamma_M$, with $\delta p$ a free parameter. The location 
of $\nu_m$ is set by the fractional electron energy $\epsmax = 0.5$ up to $\gamma_M$ 
(\eq [\ref{gmmax2}]). Smaller values of $\epsmax$, but larger than 0.1, also provide 
acceptable fits. The afterglow shown in Figure 1 is mildly relativistic after 10 days, 
so that departures from the analytically expected light-curve $F_{\nu > \nu_i} \propto 
t^{-p}$ allow the model to accommodate both the $t^{-1.2}$ optical decay before 2 days 
and the $t^{-0.8}$ fall-off of the radio emission after 10 days. As illustrated in Figure 4 
the effect of interstellar scintillation (Goodman 1997) is essential in explaining the 
departures between the observations and the model radio fluxes. 

 We note that a jet interacting with a wind-like external medium with $A_* \sim 1$ yields 
fits with acceptable $\chi^2$, but produces millimeter fluxes that exceed the $2\sigma$
upper limits shown in Figure 4 (right panel).

\subsection{Parameter Ranges and Afterglow Energetics}
\label{ranges}

 The electron index $p$ and the initial jet opening angle $\theta_0$ are determined by the 
index of the power-law emission decay and by the time when jet effects set in, respectively 
(see \eqs [\ref{tjet}], [\ref{Fs0}], and [\ref{Fs2}]). 
The remaining four model parameters, ${\cal E}_0$, $n$, $\epsel$, and $\epsmag$, can be 
determined from the three break frequencies $\nu_a$, $\nu_i$, $\nu_c$ (see \eqs [\ref{gminj}], 
[\ref{gmcool}], [\ref{nu}], and [\ref{gmabs}]), and the synchrotron flux at the peak of the 
spectrum. Finding the location of the spectral breaks requires observations spanning a wide 
frequency range, from below the lowest break (self-absorption) to above the highest break 
(cooling, more likely). Even if this requirement is satisfied, observations in only three 
domains (radio, near infrared--optical, and $X$-ray) do not determine all the spectral breaks 
unless the unconstrained break(s) cross(es) an observing frequency. This does not seem to 
be the case for the afterglows analyzed here, observations providing at most three ``strong" 
constraints for four free model parameters. The number of constraints is even smaller if 
the interval between two adjacent observational frequencies does not contain a spectral break. 
For instance, in the case where $\nu_x < \nu_c$, the $X$-ray fluxes can be predicted from 
the optical emission and the spectral slope $\beta(p)$. If consistency is found between the 
model and the observed $X$-ray fluxes, \ie if the condition $\nu_x < \nu_c$ is indeed 
satisfied, then the $X$-ray data will provide only a ``weak" constraint, as they set only a 
lower bound on the cooling break frequency.
 
 Therefore the uncertainty in the parameters of the afterglows analyzed in this work arises 
from that the number of effective observational constraints is smaller than the number of 
model parameters. To assess these uncertainties, we find sets of parameters which yield
acceptable fits to the data, for a given isotropic equivalent energy ${\cal E}_0$. 
The distributions of the parameters for which the probability of exceeding the $\chi^2$
of the data is at least 20\% are shown in Figures 5 and 6. Note that $\theta_0$ ranges from 
about $1\deg$ to $4\deg$ and that the external medium density spans four decades. Figure 6 
shows the lack of a pattern in the electron and magnetic field parameters among the four 
afterglows.  As expected, the index $p$ is well constrained for each afterglow, the light-curve 
decay being very sensitive to it. Note that, for the afterglows modeled here, $p$ does not 
have a universal value.

 Figure 7 shows the distributions of jet energy $E_0 = ({\cal E}_0/2)(1-\cos\theta_0)$ and 
of the burst $\gamma$-ray efficiency, defined as the ratio of the isotropic equivalent 
${\cal E}_\gamma$ of the energy released by the GRB in the 20 keV -- 1 MeV range to the total 
fireball energy ${\cal E}_\gamma + {\cal E}_0$. The ${\cal E}_\gamma$ was calculated from the 
reported fluences $\Phi_\gamma$ above 20 keV (Kippen \etal 1998-9) and the measured 
redshifts: ${\cal E}_\gamma \sim 10^{53}$ erg for GRB 980703, ${\cal E}_\gamma \sim 3 \times 
10^{54}$ erg for GRB 990123, ${\cal E}_\gamma \sim 2\times 10^{53}$ erg for GRB 990510, and 
${\cal E}_\gamma \sim 7 \times 10^{53}$ erg for GRB 991216. Note that, when $\theta_0$ can be 
determined from observations, the jet energies are clustered in the $10^{50}$ to $10^{51}$ erg 
range, and that the minimum efficiency for GRBs 990123 and 991216 exceeds 20\%.

\section{Conclusions}

 We model the emission of GRB afterglows in the framework of collimated, uniform, lateral 
spreading jets interacting with an external medium. The model was used to determine the initial 
jet energy $E_0$, opening angle $\theta_0$, external medium density $n$, and parameters $\epsel$ 
and $\epsmag$ quantizing the minimum random Lorentz factor of shock-accelerated electrons and 
the strength of the magnetic field, respectively, for the afterglows of GRB 980703, 990123, 
990510, and 991216. 

 As illustrated in Figures 5 and 6, the jet aperture $\theta_0$ and the index $p$ of the 
power-law distribution of electrons injected in the down-stream region are well constrained 
by observations, as the effects of collimation are seen at a time which depends strongly on 
$\theta_0$ (\eq [\ref{tjet}]), while $p$ determines the afterglow decay rate (\eqs [\ref{Fs0}], 
[\ref{Fs2}], [\ref{Fjet}]). Other model parameters -- $E_0$, $n$, and $\epsmag$ -- are less well 
determined as observations in three frequency domains (radio, optical, and $X$-ray) provide 
at most three constraints. Observations in a fourth domain, sub-millimeter, millimeter, or 
far-infrared, could help determine all the afterglow parameters, provided that the spectral 
breaks are located between adjacent observational frequencies. 

 Analysis of the available data for the 990123 and 990510 afterglows and the jet interpretation
of the break exhibited by their optical emissions rule out a wind-like profile for the medium
which decelerates the jet. However, the emission of the 980703 and 991216 afterglows can be 
accommodated by both types of external media (homogeneous or a pre-ejected wind). For 980703,
990123, and 990510, the allowed range of external densities is below $1 \cm3$, indicating that 
these bursts did not occur in hydrogen clouds.

 For those afterglows with optical light-curve breaks we find jet energies lying mostly in
the $10^{50}$--$10^{51}$ erg range (Figure 7), and jet initial half-angles below $4\deg$.  
For the afterglow of GRB 970508, the only one observed in radio, optical and $X$-ray that is 
not included in this work due to its non-standard optical and $X$-ray brightening, Frail, 
Waxman \& Kulkarni (2000) found a similar jet energy, $E_0 = 5\times 10^{50}$ erg, a much 
larger initial half-angle, $\theta_0 \sim 30\deg$, and an external density $n \sim 1 \cm3$.

 The minimum BATSE range (20 keV -- 1 MeV) efficiency of the afterglows whose optical behavior 
indicates that the ejecta was well collimated ranges from 3\% to 30\% (Figure 7). The former 
limit is within the reach of current calculations of internal shock efficiency (Kumar 1999, 
Lazzati \etal 1999, Panaitescu, Spada \& \Meszaros 1999), but the latter exceeds it. However, 
if the energy distribution within the jet aperture is far from isotropy, such that the GRBs 
we see have an energy-per-solid angle peaking in the direction toward the observer, then the 
minimum required efficiency can be significantly smaller.

\acknowledgments{ A.P. acknowledges support from the Lyman Spitzer Jr. fellowship.
   The work of P.K. is supported in part by NSF grant phy-0070928. 
   The authors commend the work of Jochen Greiner, who maintains a very 
   useful compilation of the available information on GRB afterglows at 
     {\it http://www.aip.de:8080/~jcg/grbgen.html}.  }

\clearpage

\begin{figure*}
\centerline{\psfig{figure=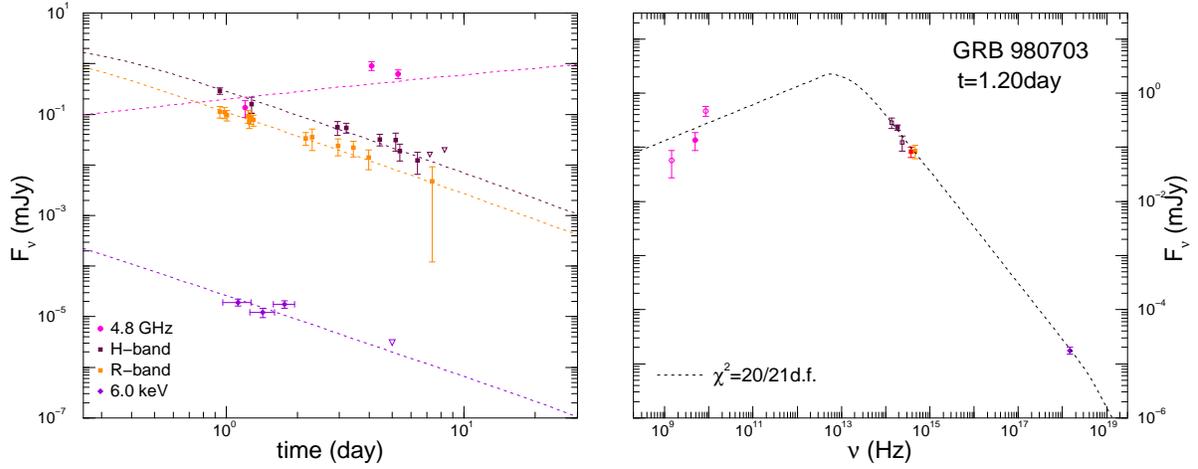,width=16cm}}
\figcaption{
  Light-curves (left panel) and $t=1.20$ day spectrum (right panel) for the afterglow 
  of GRB 980703 ($z=0.966$).
  Radio data are from Frail \etal (1999). We assume 20-40\% uncertainty 
   when no errors were reported, the larger errors being for earlier or lower 
   frequency observations, the smaller for later or higher frequencies.
  Optical fluxes are calculated from the magnitudes published by Bloom \etal (1998), 
   Castro-Tirado \etal (1999a), and Vreeswijk \etal (1999).
   The host galaxy contribution (Vreeswijk \etal 1999) is subtracted. 
   Optical fluxes are corrected for a Galactic extinction of $E(B-V)=0.061$ 
   (Bloom \etal 1998) and for the host galaxy extinction inferred by 
   Vreeswijk \etal (1999): $A_V = 1.45 \pm 0.13$. 
  $X$-ray fluxes are calculated from the 2--10 keV band fluxes reported by 
    Vreeswijk \etal (1999). 
  Triangles indicate $2\sigma$ upper limits. 
  Right panel: optical measurements closest to $t=1.20$ day are extrapolated to 
   this time using $F_\nu \propto t^{-\alpha}$, with the indices $\alpha$ reported by 
   Vreeswijk \etal (1999). Open symbols for radio fluxes are extrapolations from 
   the 4.1 and 5.3 days measurements of Frail \etal (1999) with the analytically 
   expected behavior $F_\nu \propto t^{1/2}$.
  The model shown has $\chi^2=20$ for 21 df and the following parameters:
   ${\cal E}_0=2.9\times 10^{54}\;{\rm erg}$ (isotropic equivalent energy), 
   $n=7.8\times 10^{-4} \cm3$, $\epsel= 0.075$, $\epsmag=4.6\times 10^{-4}$, $p=3.08$ .
   The initial jet angle is lower bounded by the condition that no effects of collimation 
   are seen until the later available data, which leads to $\theta_0 \simg 2.7\deg$.
   Radiative losses amount to 1.6\% at 10 days. Synchrotron emission is the main electron 
   radiative cooling mechanism at all times shown here.
}
\end{figure*}

\begin{figure*}
\centerline{\psfig{figure=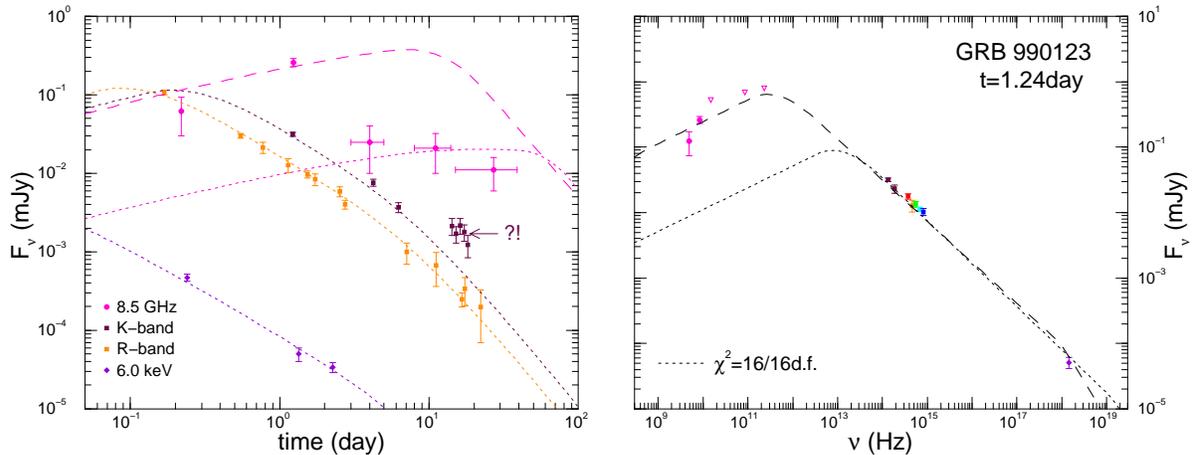,width=16cm}}
\figcaption{
  Light-curves and $t=1.24$ day spectrum for the afterglow of GRB 990123 ($z=1.61$). 
  Radio data are from Kulkarni \etal (1999b). 
  Optical fluxes are calculated from the magnitudes reported by Kulkarni \etal (1999a) 
   by subtracting the host galaxy contribution $F_r = 0.5\pm 0.1\;{\rm \mu Jy}$ 
   and $K=22.1 \pm 0.3$, and from the HST measurement reported by Odewahn \etal (1999). 
   Fluxes are corrected for a Galactic extinction of $A_r=0.04$ (Kulkarni \etal 1999a).
  $X$-ray fluxes are calculated from the 2--10 keV fluxes reported by Heise \etal (1999) 
   and Murakami \etal (1999), using the quoted spectral indices and assuming a 10\% error. 
  Other data, used for the spectrum shown in the right panel, are from 
   Galama \etal (1999) and Castro-Tirado \etal (1999b), and are extrapolated to
   $t=1.24$ day using the optical decay indices given in Castro-Tirado \etal (1999b). 
  Triangles indicate $2\sigma$ upper limits.  
  The model shown has $\chi^2=16$ for 16 df, excluding the first two radio 
   measurements and the group of late $K$-band observations (which would increase
   $\chi^2$ by $\Delta \chi^2 = 36$ for), and parameters:
   $E_0=4.6\times 10^{50}\;{\rm erg}$ (initial jet energy), $\theta_0=2.0\deg$, 
   $n=4.7\times 10^{-4} \cm3$, $\epsel=0.13$, $\epsmag=1.3\times 10^{-4}$, $p=2.32$ .
   The steepening of the optical decay is due to jet effects.
   About 9\% of the afterglow energy is radiated until 30 days. 
   Inverse Compton scatterings dominate the electron radiative cooling until several days.
  The model shown with {\it dashed} line yields a good fit to the optical and $X$-ray 
  data, but matches only the first two radio measurements, over-predicting the late
  time radio emission.
}
\end{figure*}

\begin{figure*}
\centerline{\psfig{figure=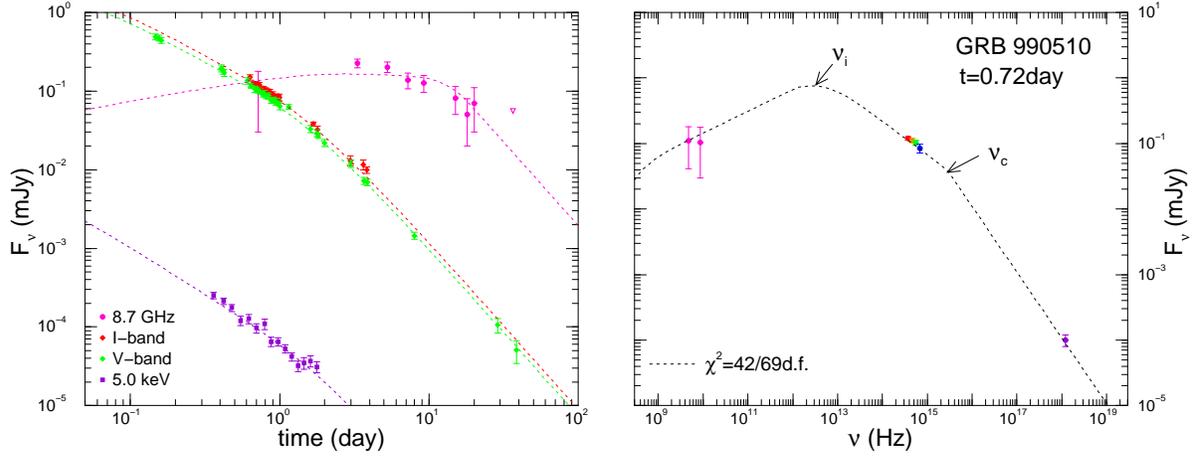,width=16cm}}
\figcaption{
  Light-curves and $t=0.72$ day spectrum for the afterglow of GRB 990510 ($z=1.62$).
  Radio data are from Harrison \etal (1999). The triangle indicates a $2\sigma$ upper limit.
  Optical fluxes are obtained from the magnitudes reported by Beuermann \etal (1999), 
   Fruchter \etal (1999), Harrison \etal (1999), Israel \etal (1999), Pietrzynski \& Udalski 
   (1999), and Stanek \etal (1999). These fluxes are corrected for dust extinction with 
   $E(B-V)=0.20$ (Harrison \etal 1999, Stanek \etal 1999). The uncertain, small contribution 
   of the host galaxy, $V_g \simg 28.0$ (Bloom \etal 2000, Fruchter \etal 2000), is ignored.
  $X$-ray fluxes are calculated from the 2--10 keV band fluxes and spectral slope published
   by Kuulkers \etal (2000).
  The model shown has $\chi^2 = 42$ for 69 df, and parameters:
   $E_0=3.0 \times 10^{50}\;{\rm erg}$, $\theta_0=2.7\deg$, $n=0.14 \cm3$, 
   $\epsel=0.046$, $\epsmag=8.6 \times 10^{-4}$, $p=2.01$. 
   The steepening of the optical decay is due to the collimation of ejecta.
   The radiative losses amount to 53\% at 40 days. At all times shown here, 
   inverse Compton scatterings is the dominant electron radiative cooling mechanism. 
}
\end{figure*}

\begin{figure*}
\centerline{\psfig{figure=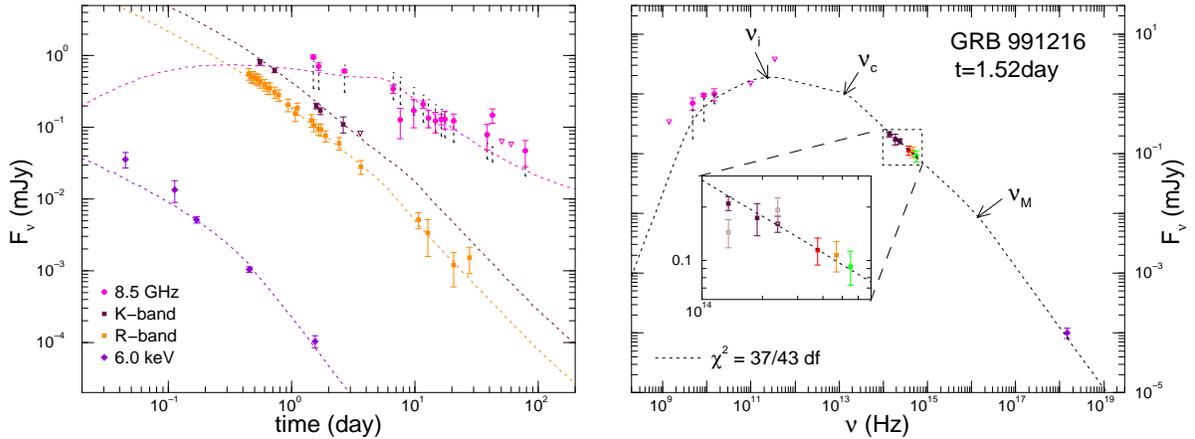,width=16cm}}
\figcaption{
  Light-curves and $t=1.52$ day spectrum for the afterglow of GRB 991216 ($z=1.02$). 
  Radio data are from Rol \etal (1999) and Frail \etal (2000).
  Optical fluxes are calculated from the magnitudes published by Garnavich \etal (2000),
   Halpern \etal (2000), Sagar \etal (2000), and Schaefer (2000). We subtracted the 
   contamination from a galaxy located near the OT, estimated by Halpern \etal (2000) 
   at $R=24.8 \pm 0.1$ (the host galaxy, identified by Vreeswijk \etal 2000, has 
   $R=26.9\pm0.4$). Optical fluxes are corrected for a Galactic extinction of $E(B-V)=0.63$ 
   (Garnavich \etal 2000, Halpern \etal 2000).
  $X$-ray fluxes are calculated from the 2--10 keV fluxes reported by Corbet \etal (1999), 
   Piro \etal (1999), and Takeshima \etal (1999).
  Right panel: the measurements closest to $t=1.52$ day are extrapolated to this time using 
  the power-law scaling found by Garnavich \etal (2000) in the near-infrared and by 
  Halpern \etal (2000) in the optical. Inset: $J$ and $K$ measurements of Halpern \etal 
  (2000) are shown with open symbols, while those reported by Garnavich \etal (2000), 
   consistent with a power-law spectrum, are indicated with filled symbols. 
  Triangles indicate $2\sigma$ upper limits. 
  The model shown has $\chi^2 = 37$ for 43 df and parameters
   $E_0=0.89\times 10^{50}\;{\rm erg}$, $\theta_0=3.4\deg$, $n=1.6 \cm3$, $\epsel=0.024$, 
   $\epsmag=0.073$, $p=1.36$, $\delta p=0.7$. The $\nu_M$ frequency is set by the fractional 
   electron energy, assumed $\epsmax=0.5$. 
  The steepening of the $X$-ray and optical fall-off is due to the passage of the $\nu_M$ break. 
  Vertical dotted lines indicate the amplitude of the interstellar 
   scintillation, which we model following the treatment given by Walker (1998).
   The afterglow radiates 82\% of its energy until 80 days. 
   The electrons cool radiatively mainly through synchrotron emission after few days.
}
\end{figure*}

\begin{figure*}
\centerline{\psfig{figure=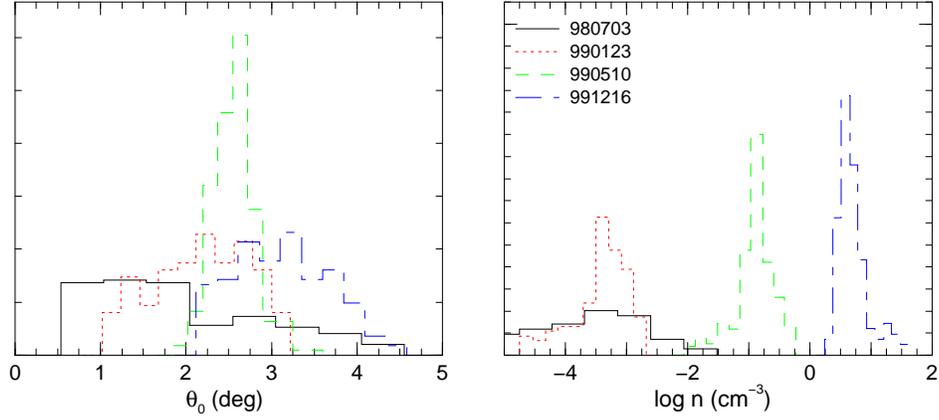}}
\figcaption{
  Distributions of the parameters $\theta_0$ (jet initial opening) and $n$ (external 
  medium density) obtained from fits to the afterglows of GRB 980703, 990123, 990510, 
  and 991216. For the models with these parameters the probability of exceeding the 
  $\chi^2$ of the data is at least 20\%. The decay of the afterglow of GRB 980703 
  does not show evidence for collimation of ejecta until several days. The initial 
  jet apertures shown in this case are obtained by requiring that $t_j = 5$ day 
  (\eq[\ref{tjet}]), therefore they are lower limits of the true $\theta_0$.
}
\end{figure*}

\begin{figure*}
\centerline{\psfig{figure=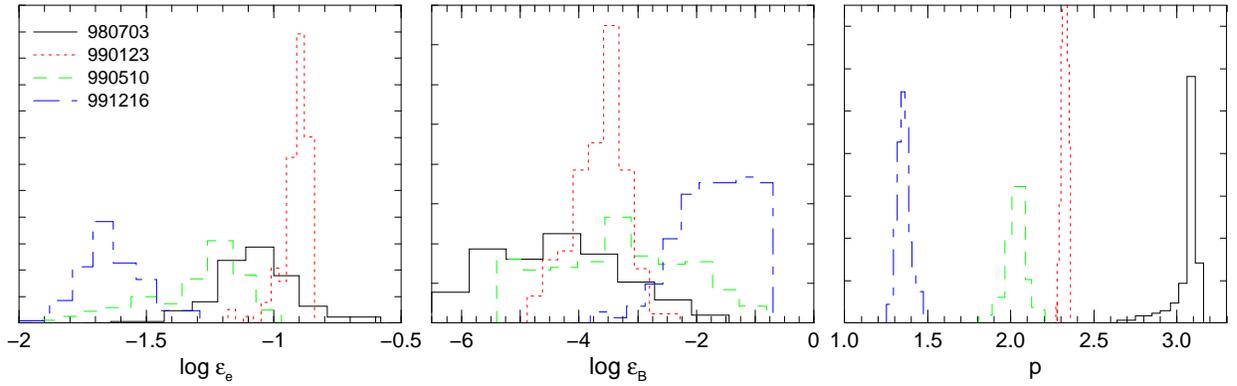}}
\figcaption{
  Distributions of the parameters $\epsel$ (for the minimum energy of the injected electrons),
  $\epsmag$ (fractional energy in magnetic field), and $p$ (exponent of the injected
  power-law distribution) for the same models as in Figure 5. Note that these four 
  afterglows have distinct allowed ranges for $p$.
}
\end{figure*}

\begin{figure*}
\centerline{\psfig{figure=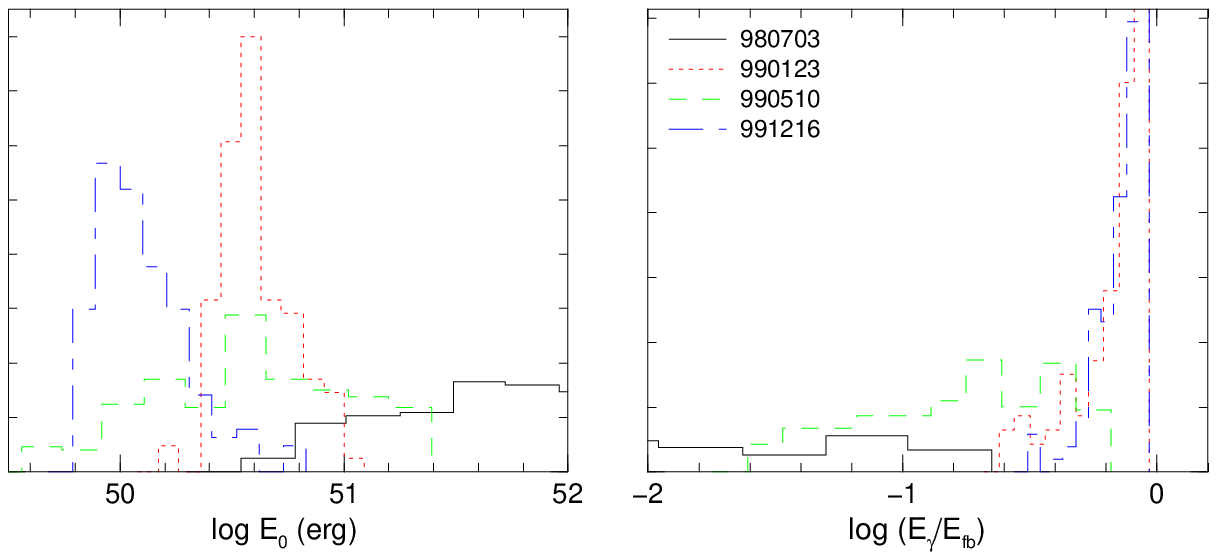}}
\figcaption{
 Distribution of the jet energy $E_0$ and GRB radiative efficiency for the models whose 
 parameters are shown in Figures 5 and 6. The radiative efficiency is the ratio of the 
 energy ${\cal E}_\gamma$ released by the GRB in the BATSE range (approximately 20 keV 
  -- 1 MeV), assuming spherical symmetry, and the initial fireball energy ${\cal E}_{fb} =
 {\cal E}_\gamma + {\cal E}_0$, where ${\cal E}_0$ is the isotropic equivalent of the 
 ejecta energy which yields acceptable fits to the afterglow emission. 
 For GRB 980703 we used the lower limits on $\theta_0$ shown in Figure 5 to calculate
 the jet energy. 
}
\end{figure*}


\begin{references}

\reference{} Beloborodov, A. 2000, ApJ, 539, L25
\reference{} Beuermann, K. \etal 1999, A\&A, 352, L26
\reference{} Bloom, J. \etal 1998, ApJ, 508, L21
\reference{} Bloom, J. \etal 2000, GCN$^*$ 756
\reference{} Campins, H., Rieke, G., \& Lebofsky, M. 1985, AJ, 90, no 5, 896 
\reference{} Cardelli, J., Clayton, G., \& Mathis, A. 1989, ApJ, 345, 245
\reference{} Castro-Tirado, A. \etal 1999a, ApJ, 511, L85
\reference{} Castro-Tirado, A. \etal 1999b, Science, 283, 2069
\reference{} Corbet, R. \etal 1999, GCN$^*$ 506
\reference{} Djorgovski, S. \etal 1998, ApJ, 508, L17
\reference{} Djorgovski, S. \etal 1999, GCN$^*$ 510
\reference{} Frail, D. \etal 1999, GCN$^*$ 141
\reference{} Frail, D., Waxman, E., \& Kulkarni, S. 2000, ApJ, 537, 191
\reference{} Frail, D. \etal 2000, ApJ, 538, L129
\reference{} Fruchter, A. \etal 1999, GCN$^*$ 386
\reference{} Fruchter, A. \etal 2000, GCN$^*$ 757
\reference{} Fukugita, M., Shimasaku, K., \& Ichikawa, T. 1995, PASP, 107, 945
\reference{} Galama, T. \etal 1998, GCN$^*$ 145
\reference{} Galama, T. \etal 1999, Nature, 398, 394
\reference{} Garnavich, P. \etal 2000, ApJ, 543, 61
\reference{} Goodman, J. 1997, New Astronomy, 2, 49
\reference{} Halpern, J. \etal 2000, ApJ, 543, 697
\reference{} Harrison, F. \etal 1999, ApJ, 523, L121
\reference{} Heise, J. \etal 1999, IAUC 7099, GCN$^*$ 202
\reference{} Huang, Y., Gou, L., Dai, Z., \& Lu, T. 2000, ApJ, 543, 90
\reference{} Israel, G. \etal 1999, A\&A, 348, L51
\reference{} Kelson, D. \etal 1999, IAUC 7096
\reference{} Kippen, R. \etal 1998-9, GCN$^*$ 143, 224, 322, 504
\reference{} Kulkarni, S. \etal 1999a, Nature, 398, 389
\reference{} Kulkarni, S. \etal 1999b, ApJ, 522, L97
\reference{} Kumar, P. 1999, ApJ, 523, L113
\reference{} Kumar, P. \& Piran, T. 2000, ApJ, 532, 286
\reference{} Kumar, P. \& Panaitescu, A. 2000, ApJ, 541, L9
\reference{} Kuulkers, E. \etal 2000, ApJ, 538, 638
\reference{} Lazzati, D., Ghisellini, G., \& Celotti, A., 1999, MNRAS 309, L13
\reference{} Mathis, J. 1990, ARA\&A, 28, 37
\reference{} \Meszaros, P. \& Rees, M.J. 1994, ApJ, 430, L93
\reference{} \Meszaros, P. \& Rees, M.J. 1999, MNRAS, 306, L39
\reference{} Murakami, T. \etal 1999, GCN$^*$ 228
\reference{} Odewahn, S. \etal 1999, GCN$^*$ 261
\reference{} Panaitescu, A., \Meszaros, P., \& Rees, M.J. 1998, ApJ, 503, 314
\reference{} Panaitescu, A., Spada, M., \& \Meszaros, P. 1999, ApJ 522, L105
\reference{} Panaitescu, A. \& Kumar, P. 2000, ApJ, 543, 66
\reference{} Pietrzynski, G. \& Udalski, A. 1999, GCN$^*$ 319, 328
\reference{} Piro, L. 1999, GCN$^*$ 500
\reference{} Rhoads, J. 1999, ApJ, 525, 737
\reference{} Rol, E. \etal 1999, GCN$^*$ 491
\reference{} Sagar, R., Mohan, V., Pandey, A., \& Castro-Tirado, A. 2000, BASI, 28, 15
\reference{} Sari, R., Piran, T., \& Narayan, R. 1998, ApJ, 497, L17
\reference{} Sari, R. \& Piran, T. 1999, ApJ, 517, L112
\reference{} Schaefer, B. 2000, GCN$^*$ 517
\reference{} Schild, R. 1977, AJ, 82, 337
\reference{} Stanek, K., Garnavich, P., Kaluzny, J., Pych, W., \& Thompson, I. \\
             1999, ApJ, 522, L39
\reference{} Takeshima, T. 1999, GCN$^*$ 478
\reference{} Vreeswijk, P. \etal 1999, ApJ, 523, 171
\reference{} Vreeswijk, P. \etal 2000, GCN$^*$ 751
\reference{} Walker, M. 1998, MNRAS, 294, 307
\reference{} $^*$ GCN Circulars can be found at {\sl http://gcn.gsfc.nasa.gov/gcn/}

\end{references}
\end{document}